\documentclass[%
reprint,
superscriptaddress,
amsmath,amssymb,
aps,
prmaterials,
]{revtex4-2}

\usepackage{graphicx}
\usepackage{dcolumn}
\usepackage{bm}
\usepackage{hyperref}

\usepackage{cleveref}
\usepackage{siunitx}
\usepackage{mathtools}
\usepackage[caption=false]{subfig}
\usepackage{enumitem}
\usepackage{multirow}
\usepackage{scalerel,amssymb}
\usepackage{verbatim}
\usepackage[dvipsnames]{xcolor}
\usepackage{lineno}
\UseRawInputEncoding

\crefname{figure}{Figure}{Figures} 
\Crefname{figure}{Figure}{Figures} 
\crefname{table}{Table}{Tables}
\Crefname{table}{Table}{Tables}
\crefname{equation}{Eq.}{Eq.}
\Crefname{equation}{Equation}{Equations}

\begin{document}
\title{Pettifor Maps of Complex Ternary Two-dimensional Transition Metal Sulphides}

\author{Andrea Silva}
\email{a.silva@soton.ac.uk}
\affiliation{Engineering and Physical Sciences, University of Southampton, UK}
\affiliation{national Centre for Advanced Tribology Study, University of Southampton, UK}

\author{Jiangming Cao}
\affiliation{Mechanical Engineering, Helmut Schmidt University, Hamburg, Germany}

\author{Tomas Polcar}
\affiliation{Engineering and Physical Sciences, University of Southampton, UK}
\affiliation{Advanced Materials Group, Department of Control Engineering, Faculty of Electrical Engineering, Czech Technical University in Prague (CTU), Czech Republic}

\author{Denis Kramer}
\email{d.kramer@hsu-hh.de}
\affiliation{Engineering and Physical Sciences, University of Southampton, UK}
\affiliation{Mechanical Engineering, Helmut Schmidt University, Hamburg, Germany}


\begin{abstract}
Alloying is an established strategy to tune the properties of bulk compounds for desired applications.
With the advent of nanotechnology, the same strategy can be applied to 2D materials for technological applications, like single-layer transistors and solid lubricants.
Here we present a systematic analysis of the phase behaviour of substitutional 2D alloys in the Transition Metal Disulphides (TMD) family.
The phase behaviour is quantified in terms of a metastability metric and benchmarked against many-body expansion of the energy landscape.
We show how the metastability metric can be directly used as starting point for setting up rational search strategies in phase space, thus allowing for targeted further computational prediction and analysis of properties.
The results presented here also constitute a useful guideline for synthesis of TMDs binary alloys via a range of synthesis techniques.
\end{abstract}

\maketitle


%
Since the discovery of graphene, 2D materials have been at the forefront of Materials Science and Discovery.
In addition to fundamental research interest~\cite{Smolenski2020}, recently their unique properties and reduced dimensionality have sparked an interest for nanoscale engineering applications.
Ideas for 2D-materials-based devices can be found in tribology~\cite{Song2018RobustSuperlub}, electronics~\cite{Das2015} and  catalysis~\cite{Pattengale2020}.
In this relatively new field, there have been so far only limited attempts to exploit the vast chemical space spanned by alloys to optimise properties.
Up to now, most research efforts have focused on identifying 2D unaries and binaries both theoretically~\cite{Mounet2018,Sorkun2020} and experimentally~\cite{Zhou2018,Shivayogimath2018}.
However, little is known about their thermodynamic phase behaviour.
The structures and ordering of possible alloys are largely unexplored territory~\cite{Domask2015}.
Only few 2D ternaries have been reported by experiments~\cite{Koepernik2016,SAEKI1987} and, while a handful of binary alloys has been studied~\cite{Gao2020,Han2020a,Chen2013}, no systematical analysis has been carried out.
But knowledge of thermodynamic behaviour is fundamental for advancing the engineering applications of 2D materials.
When properties such as bandgap and electronic transport need to be tuned to desired values by chemical doping, the presence of miscibility gaps and competing ternaries has to be taken into account~\cite{Worsdale2015a}.

The vast crystallographic and chemical spaces need not be explored by experiments alone.
Computational tools can provide guidelines to experimental synthesis, reducing the number of possible candidates by orders of magnitude.
As an example, Mounet \textit{et al.}~\cite{Mounet2018} reduced a dataset of $\SI{1e5}{}$ bulk geometries from experimental databases to 258 easy-exfoliable monolayer (ML) candidates.
As a comparison, large-scale experimental studies usually deal with dozens of candidates ~\cite{Zhou2018,Shivayogimath2018}.

In the last century, the discovery of new metallic alloys was guided by empirical methods like the Hume-Rothery rules~\cite{Abbott} and Pettiford maps~\cite{Pettifor1986}.
These rules are based on atomic proprieties like relative ionic size and electronegativity, combined through chemical intuition and experience.
The somewhat surprisingly wide validity of these simple rules in metallic alloys has been proven by experiments in the 1940s.
With the advent of Density Function Theory (DFT) and Cluster Expansion (CE) methods in the 1980s, the physics underpinning the phase diagram of metallic alloys was explored systematically, with a symbiotic relationship between experiments and simulations~\cite{Connolly1983}.
Nowadays, we are able to create large databases of materials and rationalise complex trends coupling the predictive power of DFT, the massive improvement in computation power and the availability of software tools.
These capabilities, along with experimental validation, should allow us to build on the Hume-Rothery and Pettiford rules and extend their concepts to novel classes of materials.
Indeed there are examples of such efforts in recent literature: the known empirical rules have been cast in terms of well defined probabilistic models trained on large computational datasets~\cite{Hautier2011} or extended to include the physics of oxides~\cite{Ceder2000}.

Here, a framework is presented and a dataset compiled to explore alloy possibilities for the TMD family, the most widely studied 2D material family for engineering applications.
The article is structured as follows.
The first section defines the chemical and coordination spaces considered.
Then, a metric to quantify metastability and solubility tendency in different hosts is developed.
The metric is applied to the chemical and coordination space defined in the first section, yielding the host most receptive for alloying for each transition metal (TM) pair.
In the third section, the CE formalism~\cite{Connolly1983} is used to benchmark the predictions of our metric and to identify stable orderings.
For illustration, an analysis of the phase behaviour is presented here for four representative alloys.
The Supporting Information (SI) contains further examples.
Finally, the article concludes with a discussion of how the framework could guide synthesis efforts.

\section{Chemical and Coordination Spaces}
The starting point to build the space of possible compounds is the 2D-materials database compiled by Mounet and coworkers~\cite{Mounet2018}.
The database comprises 258 mechanically stable ML structures identified from experimental bulk compounds.
Thus, the following phase stability study is conducted on ML geometries only.

\begin{figure*}[!htb]
    \centering
    \includegraphics[width=\textwidth]{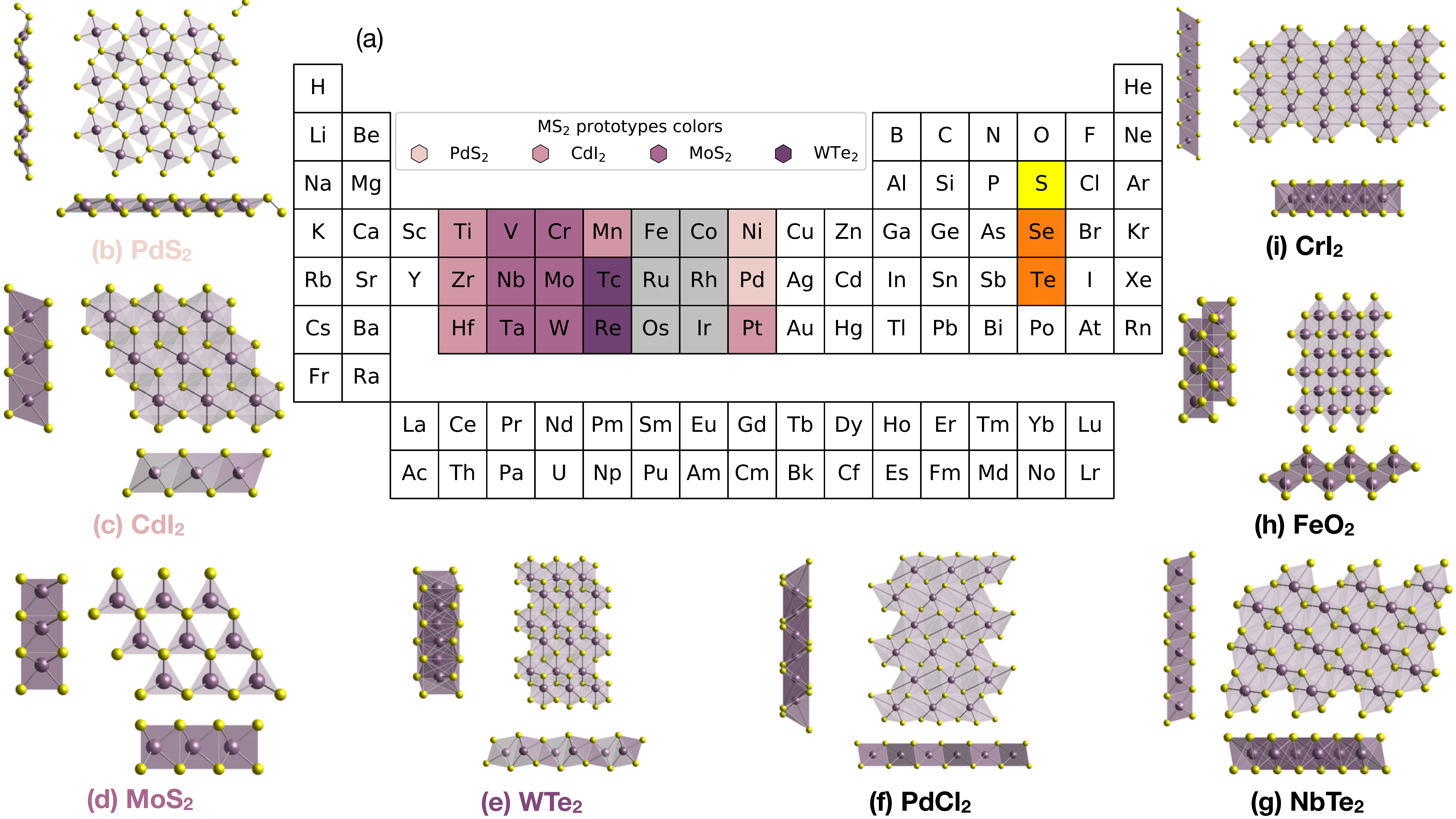}
    \caption[Periodic table with the selected TM and prototypes]{
    (a) Periodic table showing the elements selected.
    TM boxes are colored according to the MX$_2$ 2D GS prototype, as reported in \cref{fig:stab_matr}.
    Gray boxes indicate non-layered, 3D ground-state TMDs.
    Sulphur is highlighted in yellow while the other calchogenides are in orange.
    (b-i) The sides and top views of the eight MX$_2$ prototypes.
    The space group of each prototype is reported in Table SIII of the SI.
    }
    \label{fig:p_table-TM_calc}
\end{figure*}
To reduce the computational effort, the selection of the possible prototypes and elements to mix is guided by knowledge in the literature~\cite{Mounet2018,Furlan2015,Shivayogimath2018,Onofrio2017} and the original database is filtered according to the class of materials of interest.
Here, the database is scanned for compounds of the form M$_n$A$_2$, where M is a TM cation (highlighted in \cref{fig:p_table-TM_calc}a) and A is the anion, oxidising the TM (see Section I in the SI for the list of anions considered).
In selecting the prototypes, the possible cations are restricted to the transition metals considered but the anions are not limited to sulphur, as layered  prototypes that could host TMD alloys may not be expressed in terms of sulphides in the database (see Section I and table SII in the SI for details).
This search yields the eight prototypes shown in \cref{fig:p_table-TM_calc}b-i, whose space group is reported in Table SIII of the SI.
While here the symmetry of each prototype is frozen, focusing on the substitutional degree of freedom, it is in principle possible to identify pathways between these crystal structures allowing for phase transitions between the prototypes~\cite{Thomas2021ComparingGeometry}.

Intermediate TMs (Cr, Mn, Fe, Ru, Os) are considered here although they do not form layered sulphides on their own but might form ML alloys in combination with other TMs, e.g. Fe-doped MoS$_2$ ML~\cite{Furlan2015}.
Late transition metals from group XI onward are excluded, as they do not bind with chalcogenides to form layered materials~\cite{Shivayogimath2018}.
This yields the $N=21$ TMs highlighted in \cref{fig:p_table-TM_calc}a as a possible cations $M$ in the $M\mathrm{S}_2$ stochiometry.

While the methodology described here is valid for any stochiometry and cation-anion selection, our analysis will focus on $M\mathrm{S}_2$ compounds, as these are the most frequently synthesised and studied compounds of the family.
This selection yields TM $\times$ prototypes $\times$ chalcogenides = 168 binaries as a starting point for TM$_1$ $\times$ TM$_2$ $\times$ prototypes = 3528 substitutional alloys on the TM site.
The total number of candidates, although large from an experimental point of view, allows for an exhaustive theoretical analysis rather than approximate methods based on a statistical sampling of configurational space~\cite{Avery2019}.

\subsection{Lattice stability}\label{sec:formen_proto}
The total energy of each compound $M$S$_2$ in all prototypes $p$, i.e. pairs $(M,p)$, is obtained from Equation of State (EoS) calculations.
The volume range considered in the EoS is determined using the notion of covalent radius  $r_\mathrm{c}$ of the element $i$.
The protocol is described in Section II of the SI.

The energy above the ground state of each compound $M$S$_2$ in a given prototype $p$, also known as \textit{lattice stability}~\cite{Wang2004Calphad}, is given by the total energy per site with respect to the ground state (GS), i.e.
\begin{align}
\label{eq:MS2_formen}
    E_\mathrm{F}(M, p) = \frac{E(M, p)}{n} - E_\mathrm{GS}(M),
\end{align}
where $E(M, p)$ is the minimum energy of $M$S$_2$ compounds in prototype $p$ obtained from EoS calculations and $n$ is the number of sites in the metal sub-lattice, i.e. the number of TM in the unit cell.
The offset energy for each TMD $E_\mathrm{GS}(M)$ is the minimum energy across the prototype space $E_\mathrm{GS}(M) = \frac{1}{n}\min_p E(M, p)$ for layered TMD and the total energy of the 3D bulk structure $E_\mathrm{GS}(M) = \frac{1}{n}E_\mathrm{3D}(M)$ for non-layered TMDs.

The non-layered TMDs are identified by comparing the minimum-energy 2D prototype across the considered ML geometries to the GS reported in the Materials Project (MP) database ~\cite{Ong2008,Jain2013a} for the given $M\mathrm{S}_2$ compound.
The 3D geometry has lower energy than the relative 2D GS for six metal disulphide, namely FeS$_2$, CoS$_2$, RuS$_2$, RhS$_2$, OsS$_2$, IrS$_2$ (TM in gray  boxes in \cref{fig:p_table-TM_calc}).
An analysis equivalent to the one presented here but restricted to 2D geometries is reported in Section IX of the SI as it might be relevant  for experimental techniques able to bias the synthesis towards atomically thin films~\cite{Wang2021AtomicOutlook}.

For the layered TMDs, the binding energy between the layers (typically around $\SI{10}{meV/atom}$ for TMDs~\cite{Irving2017a,Levita2014a}) is neglected here, since this offset does not affect the ML phase behaviour~\cite{Silva2020a}.

\begin{figure*}[!ht]
    \centering
    \includegraphics[width=0.95\textwidth]{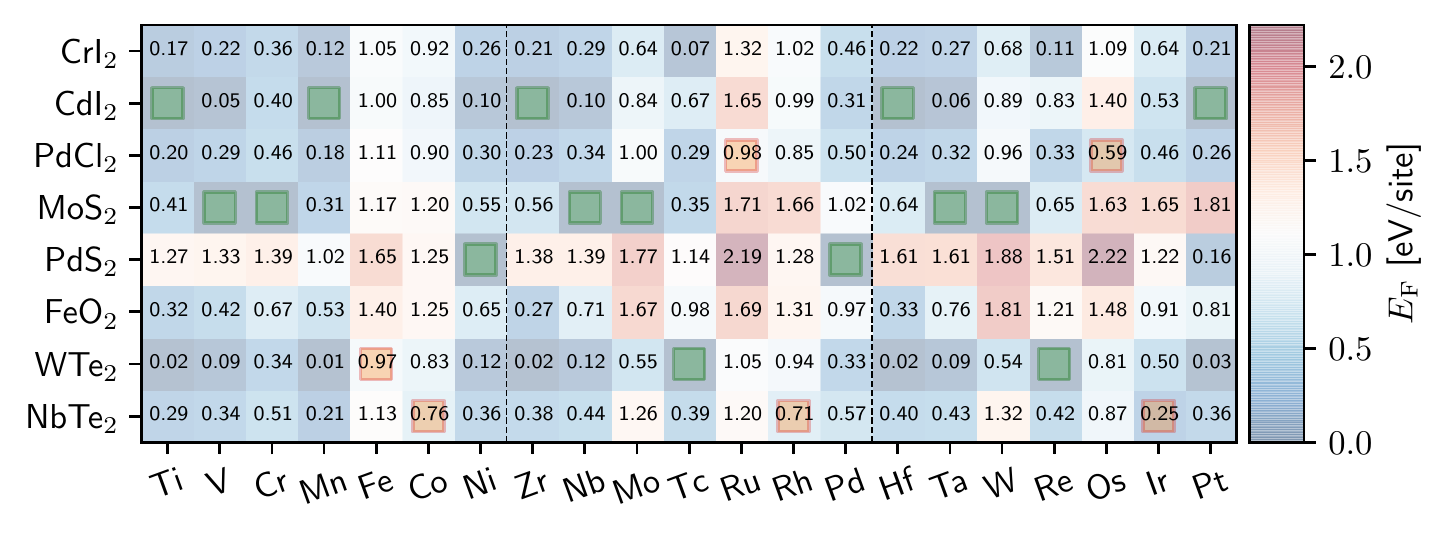}
    \caption[Lattice stability of $M$S$_2$ compound in the prototypes]{
    Lattice stability of $M$S$_2$ compound in the prototypes shown in \cref{fig:p_table-TM_calc}b-f according to \cref{eq:MS2_formen}, $M$ being one of the metals highlighted in \cref{fig:p_table-TM_calc}a.
    The colorbar on the right reports the energy above the ground state in eV over lattice sites.
    Green squares mark GS prototypes, defined by $E_F = 0$.
    Orange squares mark the lowest-energy 2D prototype of transition metals displaying a 3D GS (gray boxes in \cref{fig:p_table-TM_calc}a).
    Vertical dashed black lines separate rows of the periodic table, see
    \cref{fig:p_table-TM_calc}a.
    }
    \label{fig:stab_matr}
\end{figure*}
\Cref{fig:stab_matr} reports the energy above the ground state per lattice site defined in \cref{eq:MS2_formen} for the selection of TMs and prototypes shown in \cref{fig:p_table-TM_calc}.
Each column shows the energy above the ground state of the given TM in the eight prototypes with respect to the identified 2D GS.
Green squares mark the GS of layered TMDs and orange squares mark the lowest-energy 2D prototype of non-layered TMDs.
As a guide to the eye, each entry is colored according to its energy above the ground state, as reported by the colorbar on the right, and periodic table rows are separated by vertical dashed lines.

The ground states of known layered compounds are identified correctly according to the MP database: $d^2$-metal TMDs (TiS$_2$, ZrS$_2$ and HfS$_2$) display octahedral CdI$_2$ coordination, \cref{fig:p_table-TM_calc}c.
The MoS$_2$ prismatic prototype, \cref{fig:p_table-TM_calc}d, is the GS of $d^4$ TMDs, while the $d^{10}$ metals Ni and Pd are found to favour the square planar PdS$_2$ prototype, \cref{fig:p_table-TM_calc}b ~\cite{Ong2008,Jain2013a}.
A systematic comparison of the predicted ground state for the 21 pristine compounds in \cref{fig:stab_matr} with experimental and computational data available in the literature~\cite{Su2020HVS2,Zhuang2016HVS2,Isaacs2016HVS2,Habib2019CrS2,Zhuang2014CrS2,Nair2022FeS2,Zhang2015FeS2,Wang2020NiS2,Tang2022NiS2,Bergerhoff1987,Ong2008,Jain2013a} indicates that our protocol correctly describes the energetics of the considered chemical space.
The comparison is reported in Section III and Table SIV of the SI.

Moreover, the larger steric hindrance of heavier TMs in the same group raises the energy above the ground state of unstable prototypes.
This can be observed by following the row relative to prototype PdS$_2$ in \cref{fig:stab_matr}: $E_\mathrm{F}(\mathrm{Ti}, \mathrm{PdS}_2) = \SI{1.27}{eV/site}$,
$E_\mathrm{F}(\mathrm{Zr}, \mathrm{PdS}_2) = \SI{1.38}{eV/site}$ and
$E_\mathrm{F}(\mathrm{Hf}, \mathrm{PdS}_2) = \SI{1.61}{eV/site}$.
For prototype CdI$_2$: $E_\mathrm{F}(\mathrm{Cr}, \mathrm{CdI}_2) = \SI{0.40}{eV/site}$,
$E_\mathrm{F}(\mathrm{Mo}, \mathrm{CdI}_2) = \SI{0.84}{eV/site}$ and
$E_\mathrm{F}(\mathrm{W}, \mathrm{CdI}_2) = \SI{0.89}{eV/site}$.

Finally, it is important to appreciate the scope of validity and the possible sources of errors in the dataset presented here.
The DFT calculations performed are spin-polarised, thus non-magnetic and ferromagnetic groundstate are correctly described.
Antiferromagnetic (AFM) orderings are not considered, as calculations are performed in cells comprising a single TM site.
The only AFM orderings for the considered stoichiometry are reported for NiS$_2$ and MnS$_2$~\cite{Yu2015thermoChem}.
While important for materials properties, AFM GS in layered TMDs are usually almost degenerate in energy with FM states~\cite{Yu2015thermoChem}.
Moreover, no Hubbard correction (GGA+U) is included here.
The effect of Hubbard U on the relative total energy for the considered TMD stoichiometry is negligible~\cite{Yu2015thermoChem}, but a detailed benchmark must be carried out when applying our protocol to different stoichiometries, as discussed in the Methods section.

\section{Ideal Solid Solution Limit}
Starting from the lattice stability matrix in \cref{fig:stab_matr}, a question arises naturally: is it possible to identify which metals are likely to mix in a given prototype?
A straightforward approach to explore this question is the ideal solid solution limit, a non-interacting model based on the relative energy of pristine TMDs defined in \cref{eq:MS2_formen}.
Given a binary alloy in a prototype $p$, $M_x Q_{1-x}\mathrm{S}_2|_p$, the ideal solid solution represents a model with negligible interactions between the fraction $x$ of sites occupied by $M$ and the remaining $1-x$ sites occupied by $Q$.
In the energy-composition space, the system behaviour is represented by the line connecting the energy above the ground state of $Q\mathrm{S}_2$ in prototype $p$ at $x=0$ with the energy above the ground state for $M\mathrm{S}_2$ at $x=1$ in the same prototype, i.e. the element ($Q,p$) and ($M,p$) of the matrix in \cref{fig:stab_matr}, respectively.
Hence, in the ideal solid solution model, the energy above the ground state of a mixed configuration at concentration $x$ is given by:
\begin{equation}
\label{eq:SS_formen}
    E_{Q, M, p}^0(x) = x E_\mathrm{F}(M,p) + (1-x) E_\mathrm{F}(Q,p).
\end{equation}
By construction, this energy is exactly zero everywhere if $M$ and $Q$ share the same ground-state structure $p$, $E_\mathrm{F}(M,p) = E_\mathrm{F}(M,p) = 0$.
In any other case, the energy will be positive: suppose the metal $M$ has a ground-state geometry $p'\neq p$, the fraction $x$ of material $M\mathrm{S}_2|_p$ would transform into $p'$ to reach equilibrium at zero temperature.

The model effectively quantifies the metastability at zero temperature of alloys in a selected prototype $p$ as a function of concentration $x$.
By construction, this model cannot predict stable mixtures, i.e. negative formation energies, but can be used to estimate the likelihood of solubility and phase separation in a system: the lower the metastability of the solid solution model, the smaller any entropic or chemical stabilising mechanisms must be to stabilise alloys under synthesis conditions.
As an example, let us consider the effect of finite temperature in the solid solution model.
The equilibrium of an alloy in the prototype $p$ at temperature $T$ is determined by the free energy $F_{Q,M,p}(x,T) = E^0_{Q,M,p}(x) - TS(x)$, where the substitutional entropy of a binary alloy is a function of the concentration $x$ only, independent of the elemental pairs:
\begin{equation}
    S(x) = -[x \log x+(1-x) \log(1-x)],
\end{equation}
which counts possible configurations of the two atom types on the metal sub-lattice~\cite{ford2013statistical}.
If there exists a concentration and temperature $(x^*, T^*)$ at which $T^*S(x^*)>E^0_{Q,M,p}(x^*)$, then the free energy becomes negative and the mixture is thermodynamically stable  (see Section IV of the SI for an example).
Note the free energy $F_{Q,M,p}(x,T)$ of different hosts $p$ intersect at the same composition $x$ found for the energy above the ground state $E^0_{Q,M,p}(x)$, the entropy $S(x)$ being a function of concentration only.
Thus, the simpler linear energy model in \cref{eq:SS_formen} yields the same relative energy ordering of the prototypes, as shown in SI Section IV.
For an example of an electronic-driven stabilisation mechanism present also at zero temperature, see the discussion in the SI Section X.A.3 of the ternary GS of (Mo:Nb)S$_2$ and (Mo:Ta)S$_2$ shown in \cref{fig:metric_example}a,b.

\subsection{Metastability Metric}\label{sec:solub_metric}
\begin{figure}[!htb]
    \centering
    \includegraphics[width=\columnwidth]{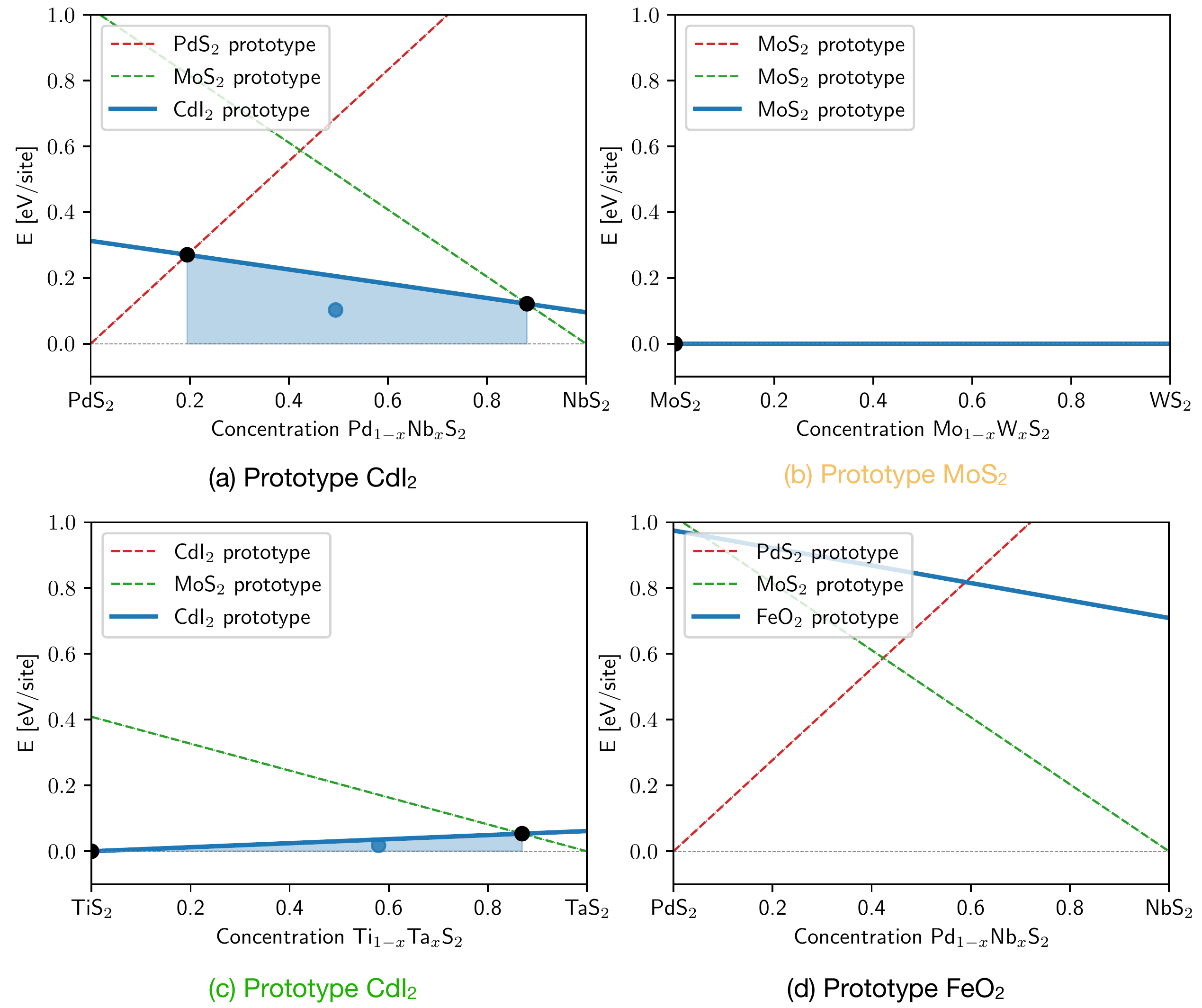}
    \caption[Examples of metastability metric]{
    Metastability metric construction (blue lines) for (a) (Pd:Nb)S$_2$ in CdI$_2$ prototype (b) (Mo:W)S$_2$ in MoS$_2$ prototype, (c) (Ti:Ta)S$_2$ in CdI$_2$ prototype, and (d) (Pd:Nb)S$_2$ in FeO$_2$ prototype.
    Blue-shaded areas highlight the extent of the metastability window in the energy above the ground state - concentration $(x, E)$ space.
    Blue circles mark the centroids of the area below the solid solution energy within the metastability window.
    Red dashed lines show the energy in the prototype of the left end-member, $x=0$.
    Green dashed lines show the energy in the prototype of the right end-member, $x=1$.
    When the considered prototype (blue line) coincides with on of the GS prototype, the line relative to the latter is hidden.
    The color of the title matches the entry highlighted in the matrix in \cref{fig:optimal_proto}.
    }
    \label{fig:metric_example}
\end{figure}

In order to make the relative metastability between prototypes quantitative, a metric in the composition-energy space is needed to compare different combinations.
Consider a prototype $p$ and two metal sulphides $M$S$_2$ and $Q$S$_2$ with GS prototype $p_M$ and $p_Q$, respectively.
The convex hull across all phases in the concentration-energy space is the line $E=0$ connecting the energies of the end-members in their respective GS prototypes, dashed gray lines in \cref{fig:metric_example}.
A point on this line at the fractional concentration $x \neq 0,1$ represents a phase separating system where the fraction $x$ of $M\mathrm{S}_2$ is in its GS prototype $p_M$ and the remaining $1-x$ is in its own GS $p_Q$.
For a configuration to be stable, its energy must be lower than this hull.
As our model by definition cannot break this hull, we characterise the metastability of a model alloy by its positive energy above the ground state, i.e. its distance from the hull~\cite{Sun2016a}.

We define a descriptor intended to capture the energetic ``disadvantage'' of a particular prototype $(p, Q, M)$ relative to the relevant binary ground states as follows.
The metastability window of the $(p, Q, M)$ triplet is defined as the range of concentration $x$ where the distance from the hull in \cref{eq:SS_formen} within the prototype $p$ is lower or equal to the distance from the hull within the ground-state prototypes $p_M$ and $p_Q$, as shown by blue regions in \cref{fig:metric_example}.

Let us apply this construction to an example: consider the energy above the ground state in the solid solution model of the (Pd:Nb)S$_2$ alloy in \cref{fig:metric_example}a.
The blue line refers to the energy above the ground state of the CdI$_2$ prototype, while the red and green dashed lines refer to the ground state of the PdS$_2$ end-member $x=0$ (PdS$_2$ prototype) and NbS$_2$ end-member $x=1$ (MoS$_2$ prototype).
The metastability of the prototypes varies as a function of the concentration.
Near the respective end-members, the ground-state prototypes are favoured, e.g. the PdS$_2$ prototype has lower distance from the hull in the range $x \in [0,0.2]$.
The CdI$_2$ prototype lies closer to the hull in range $x \in [0.2,0.9]$, suggesting that a metastable solution in this range in this prototypes is more likely than in either of the two ground-state prototypes.

When the two TMDs share the same prototype GS, the distance from the hull in that prototype is zero everywhere, like in \cref{fig:metric_example}b.
In this case the metastability window extents from 0 to 1, suggesting that solubility is likely.
When the prototype $p$ is the ground-state for one of the metals, the metastability window extends from the extremal concentration, $x=0$ or $x=1$, up to the intercept with the energy above the ground state in the other prototype, as shown in \cref{fig:metric_example}c.
Finally, a metastability window might not exist for a given triplet, as shown in \cref{fig:metric_example}d: the distance from the hull in the FeO$_2$ prototype is higher than in either ground-state prototypes for any concentration.
In this case, the formation of alloys within this prototype is unlikely.

Applying the construction depicted in \cref{fig:metric_example} to all TM pairs yields a $N \times N$ matrix, for each prototype $p$.
Each entry of these \textit{metastability matrices} are a $2 \times 2$ matrix containing the bounds of the metastability window and the energy above the ground state in \cref{eq:SS_formen} evaluated at the metastability limits, i.e. minimum and maximum hull-distance within the window.
The matrices associated with each prototype are reported in Section V and the dataset of the SI.

\subsection{Optimal Prototypes}
Given a pair of TMs, the prototype most receptive for alloying can be identified by comparing the metastability windows in different prototypes build from the metastability metric in the previous section.
A function associating a score to each metastability window needs to be defined in order to rank different prototypes.
This ranking has to assign a single value to the metastability windows of TM$_1$-TM$_2$-prototype triplets.
The following parametric function is chosen as goal function
\begin{equation}
\label{eq:fit_function}
   f_\zeta(w,\epsilon) = \zeta^2 \frac{\sqrt{w}}{\zeta^2+\epsilon^2},
\end{equation}
where $w$ is the width of the metastability window and the energy penalty $\epsilon$ is the hull-distance of the centroid defined by the window in the energy-concentration space, i.e. blue points in \cref{fig:metric_example}.
Thus, the function encourages large metastability windows $w$ and discourage large energy penalties $\epsilon$.
Details regarding the goal function and the selection of the appropriate weight $\zeta$ for the present dataset are reported in Section VI of the SI.

\begin{figure*}[!htb]
    \centering
    \includegraphics[width=\textwidth]{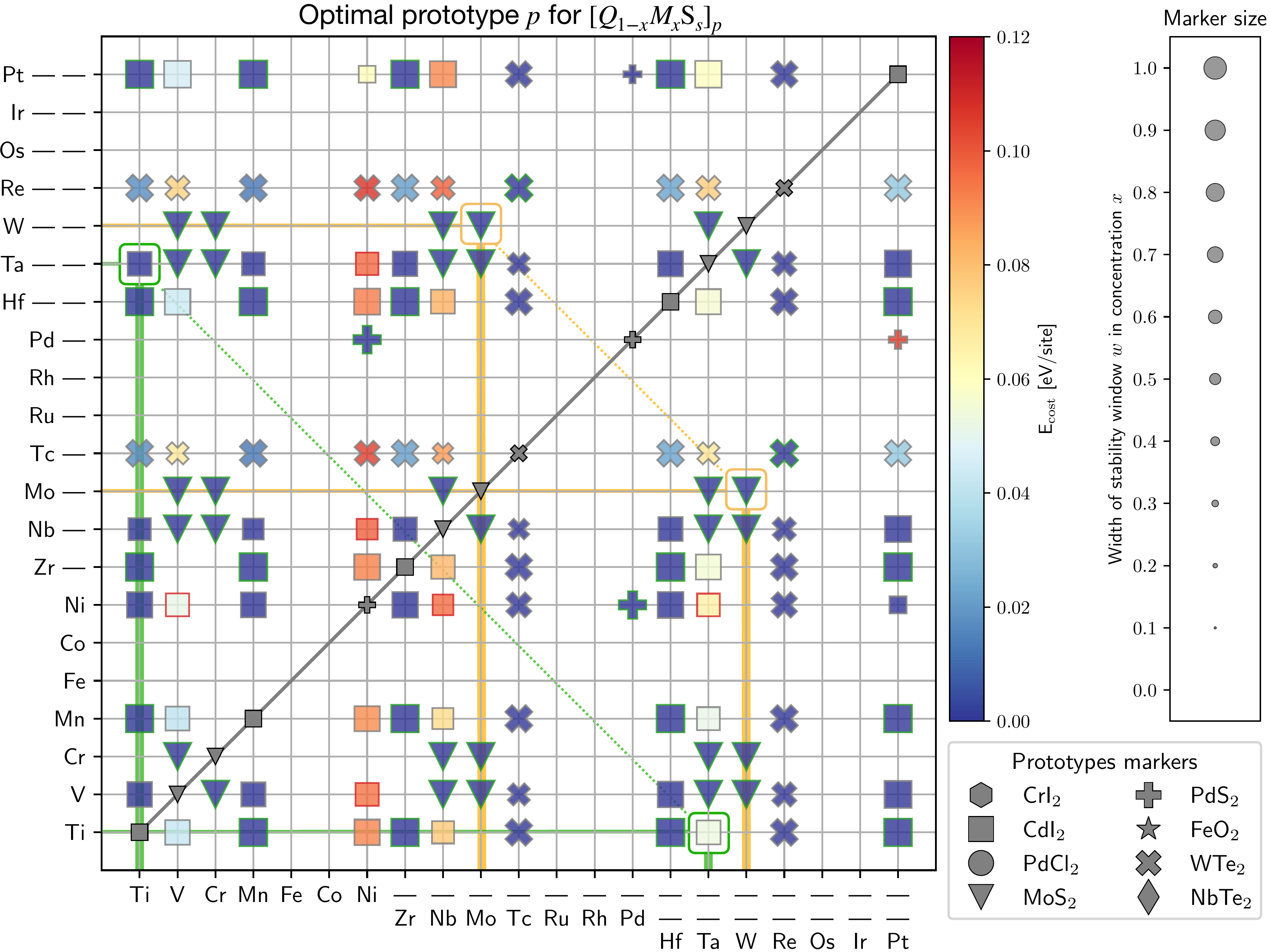}
    \caption[Optimal prototype host for TM pairs]{
    Optimal prototype for TM pairs.
    The colourcode shows the energy cost at each end of the window, in eV/site.
    The scale is reported in the first colorbar on the right.
    The energy cost of the mark refers to the end of the metastability window closer to the $M\mathrm{S}_2$ indexed by the $x$ axis.
    The size of the marker encodes the metastability window size, as reported by the legend on the far right.
    The edge color of each marker indicates whether the optimal prototype is the ground state of both (green), one (gray) or neither (red) the pristine TMDs comprising the $(M:Q)\mathrm{S}_2$ mixture.
    Marker-prototype correspondence is reported in the legend at the bottom right.
    Markers on the diagonal show the GS prototype of the corresponding metal disulphide.
    Green and orange lines highlight the entries relative to the examples shown in \cref{fig:metric_example}b,c and discussed in the main text.
    For a version without examples highlights and one without thresholds see SI Figure S13a,b.
    }
    \label{fig:optimal_proto}
\end{figure*}
The optimal prototypes for each pair of transition metals, selected by $f_\zeta$ with $\zeta = \SI{0.080}{eV/site}$, are shown in \cref{fig:optimal_proto}.
The symbol assigned to each entry refers to the optimal prototype, as shown in the lower legend; symbols on the diagonal mark the 2D-GS prototype for that transition metal.
The size of each marker shows the width of the metastability window associated with that metal pair in that prototype.
The colour code of each $Q,M$ entry shows the energy above the ground state of $Q_{1-x}M_x\mathrm{S}_2$ at each end of the metastability window, as indexed by the metal on the horizontal axis.
For example, consider the Ti$_{1-x}$Ta$_x$S$_2$ binary in the CdI$_2$ prototype, whose energy landscape is reported in \cref{fig:metric_example}c.
Follow the green lines in \cref{fig:optimal_proto} to the entry in the upper triangle, Ta row and Ti column.
This entry shows the energy above the ground state on the Ti-side of the metastability window, left-hand-side in \cref{fig:metric_example}c.
Since the CdI$_2$ prototype is the GS of TiS$_2$, the energy above the ground state on this side is zero, indicated by deep blue color.
Conversely, the entry in the lower triangle, Ti row and Ta column, shows the energy above the ground state on the Ta-side of the metastability window, right-hand-side in \cref{fig:metric_example}c.
Since the CdI$_2$ prototype is not the TaS$_2$ native prototype, the energy on this side is positive, light-blue color.

The same procedure, following the yellow lines, applies for Mo$_{1-x}$W$_x$S$_2$ binary in MoS$_2$ prototype, whose energy landscape is reported in \cref{fig:metric_example}b.
The end-members share the same GS, hence the plot shows two large, deep blue symbols with zero energy penalty.

\Cref{fig:optimal_proto} provides a visual tool to navigate the possible mixtures of transition metals within the sulphur planes.
Large blue marks in \cref{fig:optimal_proto} indicate a small energy penalty in the metastable window, and, thus, that miscibility between the two metals within the S host is likely.
For example, in the case of TiS$_2$ (GS prototype octahedral CdI$_2$) and TaS$_2$ (GS prototype prismatic MoS$_2$), \cref{fig:optimal_proto} indicates good miscibility in the CdI$_2$ prototype, that can be traced back to the relatively low energy above the ground state of TaS$_2$ in the TiS$_2$ native prototype, $E_\mathrm{F}(\mathrm{Ta}, \mathrm{CdI}_2)=\SI{0.06}{eV/site}$, see the lattice stability in \cref{fig:stab_matr}.
On the other hand, a high energy penalty and small metastable window likely results in miscibility gaps.
These likely phase-separating systems constitute the missing elements in \cref{fig:optimal_proto}.

The distinction between likely-mixing and likely-separating systems can be made more quantitative by extending the Hume-Rothery rules to our case.
Following the original rules, miscibility between transition metals within the sulphur host is expected if the lattice mismatch between the pristine compounds is less than 15 \% ~\cite{Abbott} (see SI Section VII for definition and values of the mismatch in these compounds).
Moreover, we extend the original rules using the metastability metric of the prototype.
Following the work by Sun \textit{et al.} ~\cite{Sun2016a} on metastability of inorganic crystals, we set a threshold of $E = \SI{120}{meV/site}$ as an upper limit for the  energy above the ground state of the optimal prototypes, as metastable compound within this range have been observed experimentally.
As a result, \cref{fig:optimal_proto} features “missing elements” where the optimal prototypes are unlikely to be receptive to alloying due to large lattice mismatch or high energy above the ground state.
Since experimental formation energies on these compounds are scarce, the threshold proposed here are tentative values that can easily be updated with novel experimental data.
The unfiltered matrix is reported in Section VIII of the SI.

As a first benchmark, the information in \cref{fig:optimal_proto} can be compared with alloys reported in the literature.
We focus on alloys of MoS$_2$, as many alloys for this well-known system are reported; consider the relevant column in \cref{fig:optimal_proto}, highlighted by the leftmost yellow line.
Zhou and coworkers~\cite{Zhou2018} recently reported synthesis of ML of (Nb:Mo)S$_2$,
which is shown as likely to mix in \cref{fig:optimal_proto}.
However, the same work reports a (Mo:Re)S$_2$ ML alloy, while the metastability window of this TM pair is small and high in energy ($\approx \SI{350}{meV/site}$ in \cref{fig:optimal_proto} (and Figure S13b in the SI).
Another recent work~\cite{Zhu2019d} reports the experimental characterisation of (V:Mo)S$_2$ ML, which is also a TM pair likely to mix according to our analysis.

Onofrio and coworkers~\cite{Onofrio2017} compiled a dataset of possible substitutional alloys of 1H-MoS$_2$ ML throughout most of the periodic table using DFT methods.
According to the authors' analysis, based on substitution in the smallest possible unit cell (roughly $x=0.5$), compounds based on all early TMs between group III and group VI show negative formation energy.
The authors prediction for metals of group V (V, Nb, Ta) and group VI (Cr, W) agree with our metastability metric.
In contrast, group IV elements (Ti, Zr, Hf) show a low likelihood of miscibility according to \cref{fig:optimal_proto}, while Ref.~\cite{Onofrio2017} report negative formation energies.
The case of (Mo:Ti)S$_2$ is discussed in more detail below, showing that the prediction of our metric agrees with CE analysis and available experimental data.

\subsection{Polymorphism}
The information in \cref{fig:stab_matr,fig:optimal_proto} can be coarse-grained to understand the tendency of different TMs to stabilise foreign hosts in mixtures.
Given a metal $M$, the energy cost of forming meta-stable phases as pure $M$S$_2$ is given by the columns of \cref{fig:stab_matr}, that report energy above the ground state of each $M$S$_2$ compound in the considered hosts $p$.
For example, consider the first column in  \cref{fig:stab_matr}.
TiS$_2$, whose GS is the perfectly octahedral CdI$_2$, exhibits a low energy penalty for the distorted octahedral coordination of WTe$_2$, $E_\mathrm{F}=\SI{0.02}{eV/site}$.
For MoS$_2$, whose GS is the prismatic coordination, the lowest-energy meta-stable prototype is distorted WTe$_2$ ($E_\mathrm{F}=\SI{0.55}{eV/site}$) and perfect CdI$_2$ octahedral displays a higher energy above the ground state of $E_\mathrm{F}=\SI{0.84}{eV/site}$.
The WTe$_2$ polymorph has indeed been observed experimentally~\cite{Pattengale2020} and the CdI$_2$ one has been reported in simulations of MoS$_2$ layers at high temperature~\cite{Nicolini2018a}.

\begin{figure}[!t]
    \centering
    \includegraphics[width=\columnwidth]{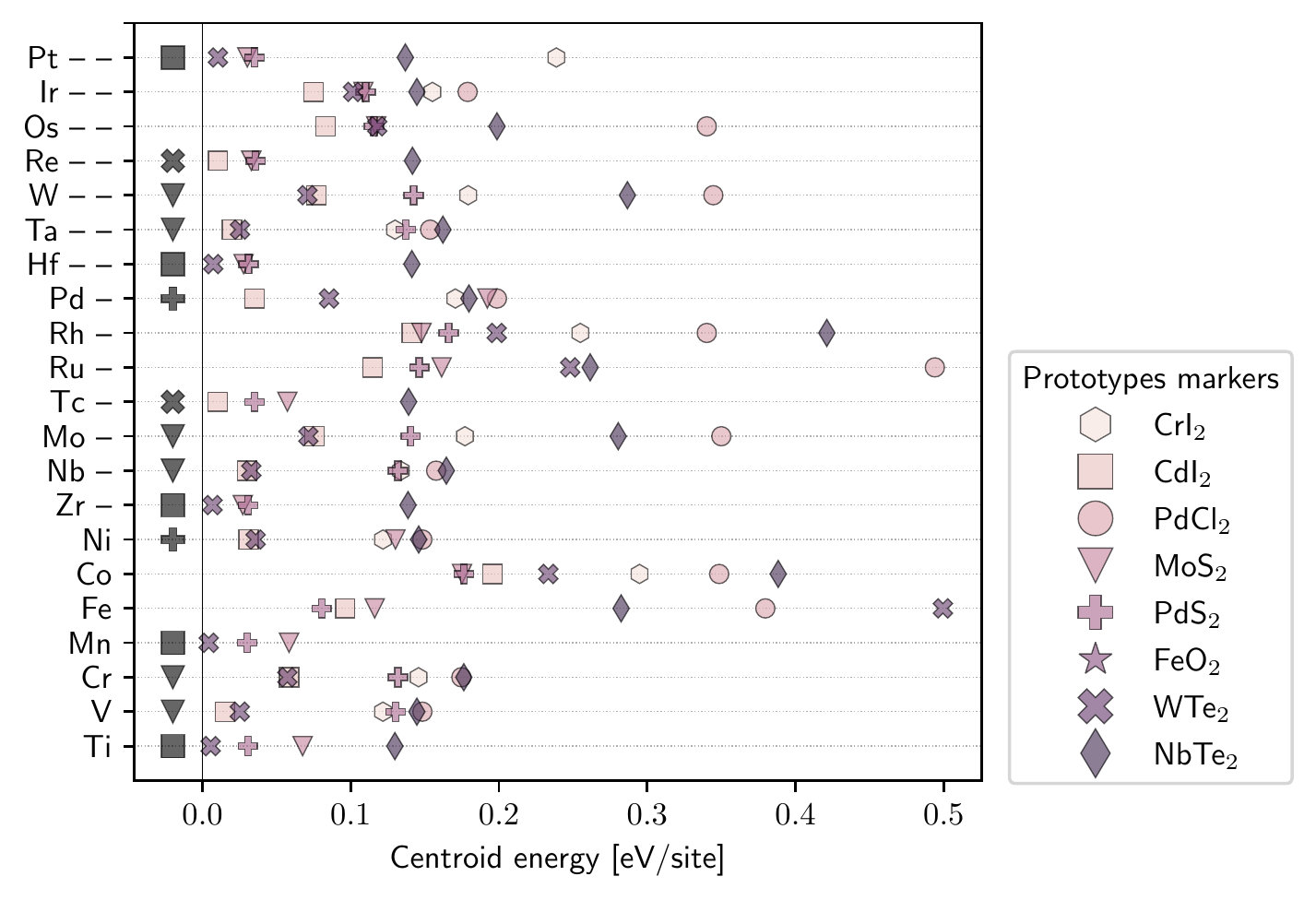}
    \caption[Polymorphism of TMDs]{
    Minimum centroid energy, $x$ axis, of all non-GS prototypes for each TM, $y$ axis.
    The legend on the right reports marker and colour associated with each prototype.
    Black marks left of $x=0$ line show the 2D GS prototype of the TM.
    }
    \label{fig:polymorphism}
\end{figure}
Similarly, the metastability metric helps to evaluate the tendency of a metal $M$ to stabilise non-native hosts when alloyed with a second metal $Q$.
Purple-shaded marks in \cref{fig:polymorphism} report the minimum centroid energy penalty $\epsilon$ across all possible combinations TM$_1$-TM$_2$-$p$, for each TM$_1$-$p$ pair.
A low centroid energy of a given prototype $p$ ($x$ axis) suggests that the considered metal $M$ ($y$ axis) could potentially stabilize this prototype when mixed with another metal in the sulphur host.
\Cref{fig:polymorphism} confirms the meta-stable tendencies highlighted in the previous paragraph.
The lowest-lying prototype for both Ti and Mo is WTe$_2$, meaning that alloys in this prototype could be stabilised by the presence of these metals.
A relatively low energy penalty for the CdI$_2$ prototype is observed in group V TMDs (VS$_2$, NbS$_2$, and TaS$_2$).
This suggests that these TMDs could be receptive for alloys in these meta-stable coordinations, alongside the native MoS$_2$ prototype.

\section{Metal Site Orderings}
\begin{figure}[!t]
    \centering
    \includegraphics[width=\columnwidth]{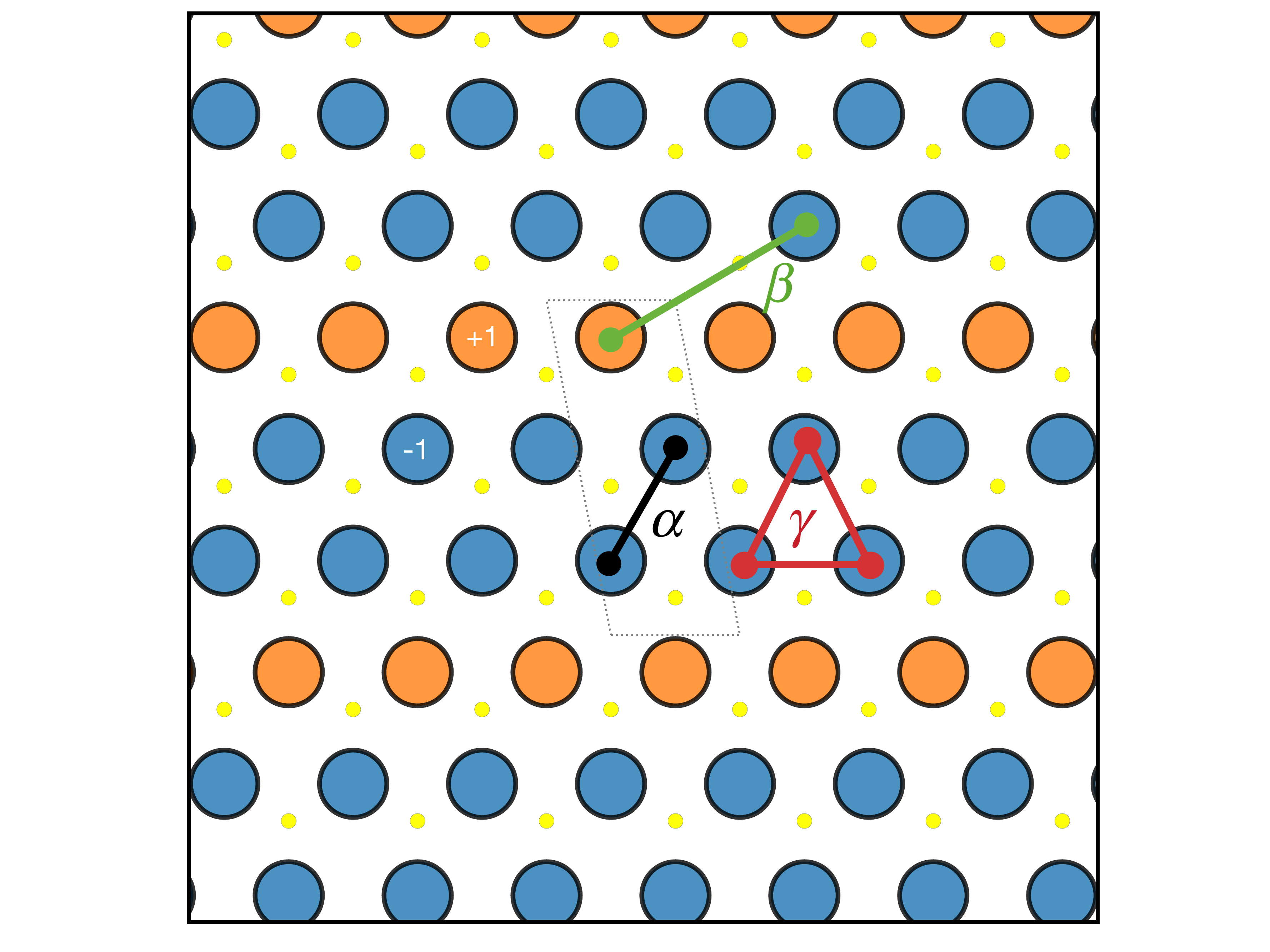}
    \caption[Ideal TMD hexagonal lattice in CE models]{
    Top view sketch of a ideal TMD hexagonal lattice, e.g. MoS$_2$ prototype, used in the CE expansion.
    The TM sub-lattice comprises of the large, black-edge circles.
    Two different species, blue and orange circles, occupy the sub-lattice.
    The occupation of each site is encoded by a two-value spin variable $\sigma_i=\pm 1$.
    The two species are here arranged in a striped pattern, whose unit cell is highlighted by grey, dashed lines.
    Small yellow circles show the spectator chalcogenide atoms.
    Colored shapes show few clusters: nearest-neighbour ($\alpha$ black line), next-nearest-neighbour ($\beta$ green line) and a triplet ($\gamma$ red triangle).
    }
    \label{fig:CE_TMD_sketch}
\end{figure}

The phase behaviour predicted by the metastability metric reported in \cref{fig:optimal_proto} can be benchmarked by exploring the stability of possible orderings and miscibility regions using a many-body expansion based on electronic-structure calculations.
The formation energy of a pseudo-binary system $M_xQ_{1-x}\mathrm{S}_2$ is modelled with the CE formalism~\cite{Connolly1983}.
The interaction between different species on the TM site sub-lattice, like the triangular one formed by orange and blue circles in \cref{fig:CE_TMD_sketch}, is modelled via a set of many-body interactions, termed clusters, e.g. the pairs $\alpha$ and $\beta$ and the triplet $\gamma$ in \cref{fig:CE_TMD_sketch}.
The sulphur atoms, yellow circles in \cref{fig:CE_TMD_sketch}, are spectators, i.e. they are considered in the DFT total energy calculations but not in the CE interaction figures.

The GS end-members are taken as reference to compute the formation energy of the ordered configuration $\sigma(x)$ at concentration $x$ in $M_x Q_{1-x} \mathrm{S}_2$:
\begin{align}
    E_{Q, M, p}(\sigma(x)) =& \left.E(\sigma(x))\right|_p  \nonumber \\
          &- x E(M, p_M) - (1-x) E(Q, p_Q), \label{eq:alloy_formen}
\end{align}
where $\left.E(\sigma(x))\right|_p$ is the total energy of the configuration $\sigma(x)$ in the host lattice defined by the prototype $p$.
$E(M, p_M)$ and $E(Q, p_Q)$ are the total energies of $M\mathrm{S}_2$ and $Q\mathrm{S}_2$ in their GS prototypes, $p_M$ and $p_Q$, respectively.
This chemical reference assures that the formation energy in \cref{eq:alloy_formen} at end-member concentration $x=0$ and $x=1$ corresponds to the energy above the ground state reported in \cref{fig:stab_matr}.

The set of geometrically distinct orderings is generated using CASM~\cite{VanderVen2010,Puchala2013,Thomas2013}.
The geometries are fully relaxed, including cell shape and volume.
The dataset is updated iteratively with stable orderings suggested by the CE model until predicted and computed convex hulls coincide. For details see the Methods section.

The following section reports our benchmark results, which cover the cases of highly-miscible TMs within the same GS host, a phase-separating system and a system with finite-miscibility of a TM in a non-native prototype.
Two other examples, one of perfect miscibility and one showing the limitation of the CE model, are presented in Sections X.C and X.D of the SI.

\begin{figure*}[!htb]
    \centering
    \includegraphics[width=0.9\textwidth]{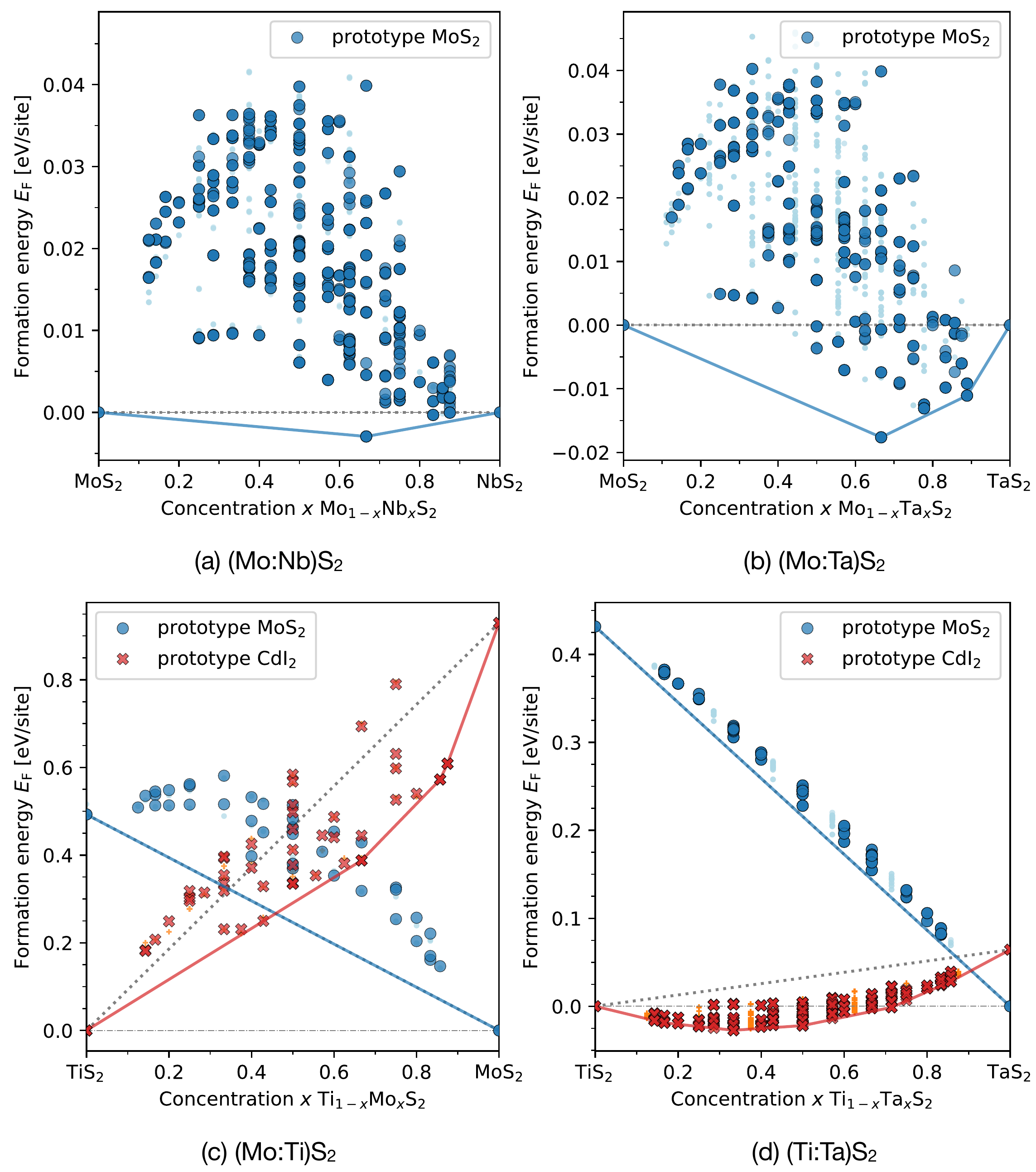}
    \caption[Formation energy of selected binary alloys]{
    Formation energies in eV/lattice site computed from DFT calculation (large black-edged symbols) and CE models (small symbols) across the whole concentration of the binary alloys reported in the title.
    Different shapes and colors refer to different prototypes as reported in the legend.
    Note that most CE energies lie behind the corresponding DFT one.
    Solid lines report the convex hull construction, marking the thermodynamic stability at fixed concentration.
    }
    \label{fig:CE_example}
\end{figure*}

\subsection{High miscibility: (Mo:group V)S$_2$ Pseudo-binary Alloys}
The metastability metric in \cref{fig:optimal_proto} predicts high miscibility for mixtures of Mo-Group V elements.
This class of alloys attracted interest as a possible realisation of MoS$_2$-based devices.
In particular, (Nb:Mo)S$_2$ alloys have been indicated as a viable p-doping solution for MoS$_2$ ML transistors~\cite{Das2015,Gao2020}.
Ta-doped MoS$_2$ composite coatings have been identified as a promising fatigue-resistant material for tribological applications~\cite{Baran2017}.

The computed alloy of both (Mo:Nb)S$_2$ and (Mo:Ta)S$_2$, reported in \cref{fig:CE_example}a,b respectively, show novel ternary GS that break the convex hull and low zero-temperature formation energy across the whole concentration range.
In particular, on the Mo-rich side (left-hand-side in \cref{fig:CE_example}a,b) substantial doping should be achievable at finite temperature, due to the absence of competing ternary ordered configurations.
On the Nb- and Ta-rich side (right-hand-side in \cref{fig:CE_example}a,b) the phase diagram is dominated by the ternary compounds breaking the convex-hull (solid lines).
These ternaries, Mo$_{1/3}$Nb$_{2/3}$S$_{2}$, Mo$_{1/3}$Ta$_{2/3}$S$_{2}$, and Mo$_{1/9}$Ta$_{8/9}$S$_{2}$, are reported here for the first time to the best of the authors knowledge.
However, the small energy scale formally stabilising these ordering at zero temperature make it likely that long range order might be destroyed at room temperature and above (see Section X.A.4 in the SI).
A good understanding of the phase behaviour of these systems is needed, especially as the doping concentration needed in p-doped devices may reach 20\%~\cite{Gao2020} and the competition with ternary phases might make synthesis problematic.

One would expect similar behaviour from Nb and Ta dopants, as the two have the same covalent radii, electronic configuration~\cite{mendeleev2014} and same lattice parameter in TMD compounds.
Indeed the qualitative behaviour is the same for both systems, as predicted by the metastability metric.
Quantitative behaviour differs slightly: a single ternary Nb$_{2/3}$Mo$_{1/3}$S$_2$ breaks the hull in the \cref{fig:CE_example}a while the Ta system displays a richer landscape with competing ternaries Ta$_{2/3}$Mo$_{1/3}$S$_2$ and Ta$_{8/9}$Mo$_{1/9}$S$_2$.
This quantitative difference arises from subtle electronic differences in the Nb and Ta ions.
Modelling these alloys present a double challenge, as one needs to capture at the same time the many-body, non-local character of phase stability and long-range elastic interactions due to lattice mismatch between NbS$_2$ or TaS$_2$ and MoS$_2$.
The CE formalism is suited to handle the first task, while the description of elasticity is problematic~\cite{VandeWalle2002}.
Since the CE expansion is performed on a complete representation of the energy landscape of the lattice model, the CE can describe small elastic displacements, at the cost of increased complexity.
Indeed, more than a hundred orbits, up to five-vertex clusters, must be included in the model to appropriately describe the convex hull in \cref{fig:CE_example}a,b, far more than for the near-commensurate (Mo:W)S$_2$ case, as reported in Section X an Table SV in the SI.
More detailed descriptions of these different contributions and of the ternary ground states are reported in Section X.A.3 in the SI.

While the system shows miscibility gaps between stochiometric GS at zero temperature, the small formation energies in the computed configurations, typically $E(\sigma(x))<k_\mathrm{B}T_\mathrm{room}=\SI{0.025}{eV}$, suggests that these miscibility gaps close below usual synthesis temperature $T_\mathrm{synth} \approx  \SI{600}{K}$ (see Section X.A.4 in the SI).

\subsection{Phase separating: (Mo:Ti)S$_2$ Pseudo-binary Alloys}
The metastability metric in \cref{fig:optimal_proto} can help identify metal pairs that would phase separate rather than form alloys in TMDs.
As an example of this behaviour, \cref{fig:CE_example}c reports the formation energy of the (Mo:Ti)S$_2$ alloys.
This system has been analysed in detail in our previous computational work in Ref.~\cite{Silva2020a} and characterised experimentally~\cite{Hsu2001}.

A high lattice stability energy of Mo in the TiS$_2$ ground-state prototype and vice versa results in a low score in the metastability metric; see the corresponding missing entry in \cref{fig:optimal_proto} (or the small light-blue triangle in the unfiltered matrix in Figure S13b in SI).
This prediction is confirmed by the CE model in \cref{fig:CE_example}c.
No configurations in the MoS$_2$ prototype (blue symbols) display lower formation energy than the solid solution limit (solid blue line).
Within the CdI$_2$ prototype, some configurations display a lower energy compared to the solid solution limit, red crosses between the solid red line and dashed gray line, respectively.
This energy gain, however, is not enough to break the inter-prototype convex hull (dash-dotted gray line at $E=0$), resulting in an overall phase separating system.
The origin of this phase behaviour lies in the different electronic structure in the local environment of the TM,  as explained in terms of crystal field levels in Ref.~\cite{Silva2020a}.
The CE model trained on DFT data have been used to estimate solubility limits in the phase space as a function of temperature, predicting low miscibility at high temperature, in line with experimental observation~\cite{Hsu2001,Silva2020a}.

\subsection{Cross-host miscibility: (Ti:Ta)S$_2$ Pseudo-binary Alloys}
\label{sec:TiTa}
Finally, we report an example of cross-host miscibility, i.e. an alloy system between two TMDs that do not share the same GS prototype.
This case is identified by combining all the information presented here.
The starting point is the polymorphism plot in \cref{fig:polymorphism}.
Group V elements (V, Nb and Ta) show low formation energy in the CdI$_2$ prototype, which is the ground state of many TMDs (see \cref{fig:stab_matr}), e.g. group IV elements (Ti, Zr and Hf).
Consulting the metastability metric in \cref{fig:optimal_proto}, possible alloying combinations of VS$_2$ and TaS$_2$ with any group VI elements stand out as promising candidates, while NbS$_2$ displays a slightly larger formation energy and can be set aside.
Taking also the mismatch into account as stated by the adapted Hume-Rothery rules, the (Ti:Ta)S$_2$ system is the most promising candidate: the mismatch for (Ti:V)S$_2$ $l_\mathrm{VS_2}/l_\mathrm{TiS_2}=0.928$ is larger than for (Ti:Ta)S$_2$  $l_\mathrm{TaS_2}/l_\mathrm{TiS_2}=0.990$ (see Section VII of the SI).
The full metastability metric construction leading to the high score of (Ti:Ta)S$_2$ in \cref{fig:optimal_proto} is also reported in \cref{fig:metric_example}c for reference.
From \cref{fig:metric_example}c, it is also clear that the MoS$_2$ prototype is unfavorable for TiS$_2$ probably resulting in phase separation between the two metals in this prototype.
This tendency is also visible in the MoS$_2$ prototype metastability matrix in Figure S3b in the SI.

We now benchmark the prediction from the metastability metric and the updated Hume-Rothery rules against actual alloy configurations from DFT.
\cref{fig:CE_example}d reports the formation energy of (Ti:Ta)S$_2$ alloys in the CdI$_2$ (red symbols) and MoS$_2$ prototypes (blue symbols).
As predicted by the metastability metric, TiS$_2$ and TaS$_2$ segregate in the MoS$_2$ prototype: no configuration lies below the solid solution limit (straight blue line).
In the CdI$_2$ prototype, native host for TiS$_2$ but not for TaS$_2$, the alloyed configurations lie below both the solid-solution line (dotted gray line) and the cross-host solid-solution hull (dash-dotted gray horizontal line) from $x \approx 0$ up to $x \approx 0.7$.
While at zero temperature only the GS on the convex hull (red solid line) are stable, the energy scale is small compared to room temperature, suggesting that solid-solution alloys in the CdI$_2$ prototype should be possible to synthesise in experiments, e.g. with CVD methods.
Indeed, there are reports of (Ti:Ta)S$_2$ solid solution alloys in the literature ~\cite{THOMPSON1972}, although no crystallography data or solubility limits are available to date.
This experimental confirmation validates the exploration approach outlined in this section.

\section{Conclusions}
We presented a systematic analysis of possible alloys in the TMD chemical space.
The metastability metric provides a simple yet useful picture to guide in-depth computational studies and experimental synthesis.
Predictions by the metastability metric are in good agreement with alloy systems reported in literature.
Moreover, many-body expansion based on electronic-structure methods of selected binary alloys confirm the predictive power of the metric both in identifying phase separating and highly miscible systems.
While this work focused on TMDs, the methodology developed here can be transferred to any stochiometry and composition.

The optimal prototype matrix and the other tools can help to identify viable alloy candidates minimising the trial-and-error attempts, speeding up the progress of nanotechnologies.
\Cref{sec:TiTa} demonstrates a possible protocol that could be followed to aid CVD synthesis of novel ML alloys to stabilise TMs in non-native local environments.

In a wider context, the framework developed here fits in the effort of making chemical intuition quantitative.
The exploration of a large dataset, easily produced with modern DFT methods, allows to rationalize trends across the periodic table and refine the known empirical rules.
In particular, attempting to transfer the Hume-Rothery rules for metallic binaries to the class of 2D TMDs seem attractive.
Here,  we propose to replace the ionic size with the lattice parameter of the $M\mathrm{S}_2$ crystal.
The rules on electron counts and electronegativity are implicitly embedded in the lattice stability differences, along with other more complex descriptors, like the $d$-band overlap and crystal field effects, as shown in the Section X.A.3 of the SI and in Ref.~\cite{Silva2020a}.
This last rule generalisation is based on the predictive power of DFT, that has been the cornerstone of Computational Material Science in the past decades.
Here we propose that miscibility is likely if the formation energy of the metastability window defined here is lower than $\SI{120}{meV/site}$.

To summarise, we presented a set of tools and ideas that will guide computational chemists and experimentalists in charting the under-explored chemical space of TMDs.

\section*{Methods}
\paragraph{First principles calculations}
The total energy calculations are carried out with the Vienna \textit{Ab Initio} Software Package (VASP)~\cite{KresseAv1996,Kresse1993,Kresse1999}, within the PAW framework for pseudo-potentials~\cite{Blochl1994}.
The generalised-gradient-approximation to DFT as parametrised by Perdew, Burke, Ernzerhof~\cite{Perdew1996} was used in this work.
The Kohn-Sham orbitals are expanded in a plane-wave basis with a cutoff of $E_{\mathrm{cutoff}}=\SI{650}{eV}$ and the BZ is sampled with a $17 \times 17 \times 1$ mesh.
The electronic density was computed self-consistently until the variation was below the threshold of $\SI{1e-6}{eV}$.
We perform spin-polarised calculation; the electronic structure can converge to non-magnetic or ferromagnetic states, as we consider only primitive unit-cells in our calculations.
The position of the ions in the unit cell were relaxed until the residual forces were below the threshold $\SI{1e-2}{eV/\AA}$.
To ensure no spurious interactions between the periodic images, a vacuum of $\SI{20}{\AA}$ was added along the $c$ axis.

Note that while error cancellation in the stoichiometric analysis carried out here makes the Hubbard U correction not necessary, Ref.~\cite{Yu2015thermoChem} shows that this becomes fundamental in modelling thermochemical reactions involving valance changes, as the reaction enthalpy of most sulphurisation reactions is not correctly described at U=0.

\paragraph{CE model training}
The fitting procedure is carried out within the CASM API~\cite{VanderVen2010,Puchala2013,Thomas2013}.
Each configuration $\sigma_i$ is weighted according to its distance from the convex hull:
\begin{equation}
    \label{eq:CASM-weight_wHull}
    w(\sigma_i) = \exp{\left(-\frac{E(\sigma_i) - E_\mathrm{hull}(x_i)}{k_\mathrm{B}\tilde{T}}\right)}
\end{equation}
where $E(\sigma_i)$ is the formation energy of the configuration $\sigma_i$, $E_\mathrm{hull}(x_i)$ is the formation energy of the convex hull at the concentration $x$ of the configuration $\sigma_i$ and $k_\mathrm{B}\tilde{T}$ is a fictitious temperature set according to the energy scale of the problem.
These weights bias the fitting towards reproducing more accurately low-energy configurations, which are the relevant ones to capture the phase behaviour of the system.
Orbits included in the CE model are selected with a genetic algorithm based on the Distributed Evolutionary Algorithm in Python (DEAP) suite ~\cite{DEAP_JMLR2012}.
A population of 100 individuals, each starting with five random-selected orbits, evolves for 20 generations.
The best 50 models are selected from five repetitions of the evolution process.
The evolution is driven by the cross-validation score of each individual, computed using the ten-split K-fold algorithm as implemented in Scikit-learn~\cite{scikit-learn}.
In order to favour low-complexity models with fewer orbits $\phi$, a penalty $p(c) = \gamma \Sigma_{c} $ is added to the cross-validation score of each individual $c$.
$\Sigma_{c}$ denotes all the cluster functions defining the model $c$, i.e. all the orbits $\phi$ associated with non-null effective cluster interaction $J$.
A value $\gamma = \SI{1e-6}{}$ has been found to yield a good compromise between reducing the number of orbits in the selected models and retaining satisfying accuracy.
%
\section*{Data and Code Availability Statements}
Metastability metric data and the code used to generated it are included in this published article as supplementary information files as JSON database and NumPy binary files for the former and Jupyter notebooks/Python3 scripts for the latter.
CE data and code used in this study are available from the corresponding author on reasonable request.

\begin{acknowledgments}
This project has received funding from the European Union's Horizon2020 research and innovation programme under grant agreement No. 721642: SOLUTION.
The authors acknowledge the use of the IRIDIS High Performance Computing Facility, and associated support services at the University of Southampton, in the completion of this work.
DK and JC acknowledges support form the Centre for Digitalisation and Technology Research of the German Armed Forced (DTEC.Bw).
TP acknowledges support of the project CAAS CZ.02.1.01\/0.0\/0.0\/16\_019\/0000778.
\end{acknowledgments}

\section*{Author contributions statement}
A.S. and J.C. performed the simulations. A.S. and D.K. conceptualized the study and wrote the manuscripts. D.K. and T.P. supervised the work. All authors reviewed the manuscript.

\section*{Competing Interests}
The authors declare no competing interests



%


\end{document}